\documentclass[twocolumn,showpacs]
{revtex4}

\usepackage{latexsym}
\usepackage{bm}

\def\Tr{{\rm Tr}}
\def\ket#1{\mid~\!\!\!{#1}~\!\!\rangle}
\def\bra#1{\langle~\!\!{#1}~\!\!\!\mid}
\def\cR{{\cal R}}

\begin{document}

\title{On the meaning of entanglement
in quantum measurement}

\author{Fedor Herbut}
\affiliation {Serbian Academy of Sciences
and Arts, Knez Mihajlova 35, 11000
Belgrade, Serbia}

\email{fedorh@infosky.net}

\date{\today}

\begin{abstract}
Measurement interaction between a
measured object and a measuring
instrument, if both are initially in a
pure state, produces a (final) bipartite
entangled state vector $\ket{\Psi
}_{12}^f$. The quasi-classical part of
the correlations in it is connected with
transmission of information in the
measurement. But, prior to "reading" the
instrument, there is also purely quantum
entanglement in $\ket{\Psi }_{12}^f$. It
is shown that in repeatable measurement
quantitatively the entanglement equals
the amount of incompatibility between the
measured observable $A_1$ and $\ket{\Psi
}_{12}^f$; and it also equals the amount
of incompatibility of $A_1$ and the
initial state of the object.
\end{abstract}

 \maketitle

\rm The meaning of entanglement in
measurement is clarified under the
following restrictions: (i) the initial
state of the object is a state vector
$\ket{\psi }$ (with or without index $1$
depending on the context); (ii) the
initial state of the measuring instrument
(MI) is a state vector $\ket{\phi }_2^0$;
(iii) the measurement is repeatable.

The restriction to $\ket{\psi }_1$ is
primarily motivated by easy feasibility
of finding an answer. Besides, it is
natural to tackle the easier case of a
pure state before tackling that of a
mixed one. Confining oneself to
$\ket{\phi }_2^0$ is often done. It is
not very restrictive because, by
so-called purification, one can, at least
in principle, replace any initial mixed
state $\rho_2^0$ of the MI by a (more
composite) state vector $\ket{\phi
}_{23}^0$ the reduced density operator of
which equals the given mixed state:
$$\Tr_3(\ket{\phi }_{23}^0\bra{\phi
}_{23}^0)=\rho_2^0.$$

Finally, repeatable measurements have the
physical advantage over the nonrepeatable
ones that consists in the possibility to
check the result obtained on an
individual object. This makes them the
more important ones. Needless to say that
they contain as a special case the kind
of measurement that is most used in the
textbook and expert literature: ideal
measurement.

It was shown in a recent article by
Vedral \cite{VedralPRL} that {\it
entanglement} does not take part in the
transmission of information from object
to MI. One wonders if it has anything to
do with the measurement. In the present
note Vedral's mentioned result is
confirmed, and, more importantly, the
connection of the entanglement with the
measurement is clarified (at least for
the mentioned restricted case).

Let $\rho$ be an arbitrary quantum state
(density operator) and
$$A=\sum_ka_kP_k+\sum_la_lP_l\eqno{(1)}$$
an arbitrary discrete observable
(Hermitian operator) in spectral form
(with distinct eigenvalues). The
observable need not be complete, i. e.,
the eigenvalues may be degenerate. The
index $k$ enumerates the detectable
(positive probability) eigenvalues in
$\rho$. As far as this state is
concerned, $A$ can be replaced by the
first sum on the RHS of (1).

To define measurement of $A$ in $\rho$,
let an MI be given - we denote it as
subsystem $2$ -  with an initial state
vector $\ket{\phi }^0_2$ and a
composite-system unitary measurement
evolution operator $U_{12}$ "containing"
the suitable interaction with the
measured object - now subsystem $1$. The
final bipartite state is
$$\ket{\Psi }_{12}^f\equiv U_{12}
(\ket{\psi }_1\otimes \ket{\phi }_2^0)
\eqno{(2)}$$ Finally, let a so-called
pointer observable $B_2\equiv
\sum_kb_kQ_2^k+\sum_lb_lQ_2^l\enskip$ be
given (in a form analogous to that of
$A$) with, possibly, degenerate $b_k$,
and $b_l$ being zero-probability in
$\ket{\Psi }_{12}^f$.

The standard {\it definition of
measurement} of $A$ in $\rho$ requires
that the so-called {\it probability
reproducibility condition} (PRC) be valid
\cite{BLM}. It states that the predicted
probabilities $p_k$ are reproduced in the
results read on the MI:
$$\forall k:\quad p_k\equiv \bra{\psi }
P_k\ket{\psi }=\bra{\Psi }_{12}^f
Q_2^k\ket{\Psi }_{12}^f. \eqno{(3)}$$
(For brevity, $Q_2^k$ stands for
$(1\otimes Q_2^k)$.)

A measurement of A is called {\it
repeatable} if the measurement result can
be confirmed by repeating the same
measurement on the individual object
(immediately after). Let me make this
more precise.

{\bf Definition.} A {\it measurement} of
$A$ in $\rho$ (cf (2) and (3)) is {\it
repeatable} if, after "reading the
pointer" (by which one means an ideal
measurement of the pointer observable)
with the result $b_k$, the measured
object is left in a state in which the
corresponding value $a_k$ of $A$ is
predicted with {\it certainty}.

Let us write the measurement in terms of
so-called {\it state transformers}
$\{A_k:\forall k\}$, linear operators
satisfying $\sum_kA_k^{\dagger}A_k=1$ and
$\forall k:\enskip A_k^{\dagger}A_k=P_k$
(projector valued measure)
\cite{Lahti95}, \cite{Nielsenbook},
\cite{VedralRMP}. (Synonyms for "state
transformer" are "measurement operator",
"Kraus operator" etc.) The state
transformers give the change of state in
measurement: $$\ket{\psi }\enskip
\rightarrow \enskip \ket{\psi }_k^f\equiv
p_k^{(-1/2)} A_k \ket{\psi }.\eqno{(4)}$$

The state transformers are the {\it first
Kraus representation of measurement}
\cite{Kraus}. The measurement evolution
(2) is {\it the second Kraus
representation of measurement}
\cite{Kraus}. It is {\it connected} with
the first one through determining the
state in which the object is left when
the MI is "read":
$$\forall k:\quad A_1^k\ket{\psi }_1
\bra{\psi }_1(A_1^k)^{\dagger}
=\Tr_2(Q_2^k \ket{\Psi }_{12}^f\bra{\Psi
}_{12}^fQ_2^k).\eqno{(5)}$$

{\bf Lemma 1.} If the measurement of an
observable $A$ is given in terms of state
transformers $\{A_k:\forall k\}$, then
{\it a necessary and sufficient condition
for repeatability} is $$\forall k:\quad
A_k= P_kA_k.\eqno{(6)}$$ (cf (1)).

(Since the operator relation (6) is
equivalent to (6) applied to an arbitrary
vector, the claim of Lemma 1 is obvious
in view of Remark 1 below.)

{\bf Theorem 1.} Let the final state in
the measurement evolution for {\it
repeatable} measurement be the state
vector $\ket{\Psi }_{12}^f$ (cf (2)). It
can be written in the (biorthogonal)
Schmidt canonical form \cite{Peres},
\cite{FHMVdistant} in terms of (tensor
factor) state vectors:
$$\ket{\Psi }_{12}^f=
\sum_k(p_k^{1/2}\ket{\psi }_1^{fk}\otimes
\ket{\phi }_2^k)\eqno{(7)}$$ where $p_k$
are the probabilities (given by (3)),
$\ket{\psi }_1^{fk}$ are determined by
(4) (with obvious necessary modification
of the notation), and
$$\forall k:\quad \ket{\phi }_2^k\equiv
(p_k)^{-1/2} \bra{\psi }_1^{fk}\ket{\Psi
}_{12}^f \eqno{(8)}$$ (partial scalar
product). Further, $$\forall k:\quad
\ket{\psi }_1^{fk}=P_1^k\ket{\psi
}_1^{fk}\eqno{(9)}$$ and $$\forall
k:\quad \ket{\phi }_2^k=Q_2^k\ket{\phi
}_2^k.\eqno{(10)}$$

{\bf Remark 1.} The physical meaning of
(9) and (10) is that the state vectors at
issue predict definite values of the
measured observable and of the pointer
observable respectively. Namely, if
$\ket{a}$ is a state vector and $E$ an
event (projector), then the general
equivalence
$$\bra{a}E\ket{a}=1\enskip
\Leftrightarrow E\ket{a}=\ket{a}$$ is
tantamount to
$$\bra{a}E^{\perp}\ket{a}=0\enskip
\Leftrightarrow E^{\perp}\ket{a}=0$$
where $E^{\perp}\equiv 1-E$. (The latter
equivalence is obvious due to idempotency
of the projector and positive
definiteness of the scalar product.)

{\it Proof} of Theorem 1. One can write
$$\ket{\Psi}_{12}^f=\sum_kQ_2^k\ket{\Psi}_{12}^f.
\eqno{(11)}$$ (Note that on account of
the PRC definition of measurement,
$\sum_kQ_2^k$ is a ceratin event in
$\ket{\Psi}_{12}^f$.)

It is seen from (5) that the
first-subsystem reduced density operator
of $p_k^{-1/2}Q_2^k\ket{\Psi}_{12}^f$ is
a ray projector. Hence there cannot be
correlations in the terms of (11), i.e.,
in view of (5) and (4), (11) can be
rewritten in terms of state vectors as in
(7). Relation (8) is an obvious
consequence of (7), (9) follows from (6)
and (4), finally, (10) is evident from
(11).

Thus, expansion (7) is biorthogonal and
it is in terms of state vectors and
positive expansion coefficients.
Therefore, it is a Schmidt canonical form
as claimed. \hfill $\Box$

{\bf Remark 2.} Defining the subsystem
states (reduced density operators)
$$\rho_s^f\equiv \Tr_{s'}(\ket{\Psi }_{12}^f
\bra{\Psi }_{12}^f)\quad s,s'=1,2\quad
s\not= s',$$ and evaluating them from
(7), one obtains $$\rho_1^f=\sum_kp_k
\ket{\psi }_1^{fk}\bra{\psi }_1^{fk},
\eqno{(12a)}$$ $$\rho_2^f=\sum_kp_k
\ket{\phi }_2^k\bra{\phi }_2^k.
\eqno{(12a)}$$

Comparing relations (1) and (9), and
relation (10) with the spectral form of
$B_2$, one comes to the conclusion that
the characteristic sub-bases of
$\rho_s^f$, $s=1,2$, appearing in (12a)
and (12b) respectively, are
simultaneously also characteristic
sub-bases of $A_1$ and $B_2$
respectively. Denoting the reducees in
the corresponding ranges of $\rho_s^f$,
$s=1,2$, by prim, one has
$$A'_1=\sum_ka_k\ket{\psi }_1^{fk}
\bra{\psi }_1^{fk}\eqno{(13a)}$$ and
$$B'_2=\sum_kb_k\ket{\phi
}_2^k\bra{\phi }_2^k.\eqno{(13b)}$$ Thus,
$A_1$ is complete in the range
$\cR(\rho_1^f)$ and $B_2$ is complete in
$\cR(\rho_2^f)$ though both may have been
incomplete {\it a priori}.

{\bf Remark 3.} Let us denote by $S_m$
the von Neumann entropy $S(\rho_m)\equiv
-\Tr\rho_mlog\rho_m$, $m=1,2,12.$ It is
well known that every bipartite pure
state has equal subsystem entropies
$S_1=S_2\equiv S$; zero total entropy
$S_{12}$; and its von Neumann mutual
information $I_{12}$, which is in general
defined as $S_1+S_2-S_{12}$, equals $2S$.
It is also known that $I_{12}$ is the sum
of the amount of quasi-classical
correlations equalling $S$ and the amount
of entanglement also equalling $S$
\cite{VedralHend}, \cite{Ved'},
\cite{ZurekOliv}. In our case of
$\ket{\Psi }_{12}^f$, one can see from
(7) that $S=H(p_k)\equiv
-\sum_kp_klogp_k$, which is the Shannon
entropy of the probability distribution
$\{p_k:\forall k\}$. It is also the
amount of information on the observable
$A$ contained in $\ket{\psi }$.

The {\it physical interpretation} of the
{\it amount of quasi-classical
correlations} $S\enskip
\Big(=H(p_k)\Big)$ is, in our case,
straightforward. It is, as mentioned, the
amount of information that the initial
state $\ket{\psi }$ contains on the
measured observable $A$. It is also the
amount of information that
$\ket{\Psi}_{12}^f$ contains about $A_1$,
and the same goes for $\rho_1^f$.
Further, it is the amount of information
that $\rho_2^f$ contains about $B_2$.
Besides, a bijection is established
between the characteristic values of
$A_1$ and $B_2$ in the Schmidt canonical
form (7). This is a perfect, i. e., in
the language of Shannon, a lossless and
noiseless information channel. Hence, it
is correct to interpret this quantity as
the amount of {\it information}
transmitted from object to MI (cf also
\cite{VedralPRL}).

The {\it physical interpretation} of the
{\it amount of entanglement} in
$\ket{\Psi }_{12}^f$ is not so easy.
(This is not surprising because
entanglement is a purely quantum
concept.) Such an interpretation was
given in a previous article for a general
bipartite pure state \cite{FH02}. But let
me, first, argue the case of $\ket{\Psi
}_{12}^f$ independently.

In QM an observable $A$ and a state
$\rho$ can be {\it incompatible} $[A,\rho
]\not= 0$. The physical meaning of this
relation is that $\rho$ cannot be written
as a mixture of states each having a
sharp value of $A$ \cite{FH69}. This,
like entanglement, {\it has no classical
counterpart}.

The {\it amount of incompatibility}
$E_C(A,\rho )$, and it coincides with the
amount of coherence of $A$ in $\rho$
(hence the index $C$), was called
incompatibility (or coherence) entropy,
and it was defined as {\it the entropy
increase} of the state in ideal
measurement of $A$ in it:
$$E_C(A,\rho )\equiv S\Big(\sum_kP_k\rho
P_k\Big)-S(\rho )\eqno{(14)}$$ (cf
relation (19) in ref. [12]). The formula
of L\"{u}ders \cite{Mess}, \cite{Lud} was
used for change of state in ideal
measurement.

{\bf Theorem 2.} The {\it amount of
entanglement} $S$ in $\ket{\Psi }_{12}^f$
equals the {\it amount of incompatibility
or coherence} of the measured observable
$A_1$ in $\ket{\Psi }_{12}^f$.

{\it Proof.} Definition (14) implies
$E_C(A_1,\ket{\Psi
}_{12}^f)=S\Big[\sum_k\Big(
P_1^k\ket{\Psi }_{12}^f\bra{\Psi }_{12}^f
P_1^k\Big)\Big]$. Applying the mixing
property \cite{Wehrl} to the entropy of
the orthogonal decomposition, one further
has $E_C(A_1,\ket{\Psi }_{12}^f)=H(p_k)$
(cf Remark 3).\hfill $\Box$

In theorem 2 a special case of the
general claim in ref. [12] is presented.
Namely, confining the composite-system
state space to $\cR(\rho_1^f)\otimes
\cR(\rho_2^f)$, the observables $A_1$ and
$B_2$ are replaced by their respective
reducees $A'_1$ and $B'_2$ given by (13a)
and (13b) respectively. These are
complete so-called twin observables (cf
\cite{ FH02}), and, according to the
general claim in this reference, any of
them "carries" the amount of the
entanglement in the bipartite state
vector via their entropy of
incompatibility with the state vector.

Every bipartite state vector has a
Schmidt canonical form and an
accompanying pair of complete twin
observables (cf (14) and (15a,b) in ref.
[12]), but they need not be physically
relevant. In the case at issue, they are,
because the twin observables are the two
observables $A_1$ and $B_2$, which play a
basic role in measurement.

As a consequence of the PRC definition of
measurement, one can say even more of the
amount of entanglement in
$\ket{\Psi}_{12}^f$.

{\bf Theorem 3.} The incompatibility
entropy $E_C(A_1,\ket{\Psi }_{12}^f)$, i.
e., the amount of entanglement in
$\ket{\Psi }_{12}^f$, equals the
incompatibility entropy of the measured
observable $A$ in the initial state
$\ket{\psi }$ of the object.

{\it Proof.} Direct evaluation gives
$$E_C(A,\ket{\psi })=S\Big[\sum_k(P_k \ket{\psi
} \bra{\psi }P_k)\Big]=H(p_k).$$ Besides
(14) also the mixing property of entropy
\cite{Wehrl} was made use of.\hfill
$\Box$

{\bf Remark 4.} The spectral forms (12a)
and (12b) in conjunction with (13a) and
(13b) show that the measured observable
$A_1$ is compatible with the final state
of the object: $[A_1,\rho_1^f]=0$.
Obviously, $A_1$ is compatible also with
$\rho_2^f$. The pointer observable $B_2$
is in the symmetric situation. Hence,
{\it the incompatibility} of $A_1$ (or
$B_2$) with $\ket{\Psi }_{12}^f\bra{\Psi
}_{12}^f$ is, actually, incompatibility
{\it with the correlations} in the
bipartite state.

Before "reading" the pointer, $\rho_1^f$
is the final state of the object (in the
measurement evolution). The
incompatibility of $A_1$ and the initial
state $\ket{\psi }_1\bra{\psi }_1$ has
disappeared, viz., $A_1$ and $\rho_1^f$
are compatible. But incompatibility {\it
in the same amount} appears between $A_1$
and $\ket{\Psi }_{12}^f\bra{\Psi
}_{12}^f$ or rather between $A_1$ and the
correlations in the latter. For a more
complete understanding of what is going
on with the incompatibilities, one must
take into account the "reading" of the
pointer.

{\bf Lemma 2.} "Reading of the pointer"
in the state $\ket{\Psi }_{12}^f$, i. e.,
ideal measurement of $B_2$ in this state,
produces the tripartite final state
$$\ket{\Phi }_{123}^{ff}=\sum_k\Big[p_k^{1/2}
(\ket {\psi }_1^{fk}\otimes \ket{\phi
}_2^k\otimes \ket{\phi }_3^k)\Big]
\eqno{(15)}$$ where the third factor
state vectors are eigenvectors of a
second pointer observable.

{\it Proof.} On account of theorem 1, we
can write $\ket{\Phi }_{123}^{ff}$
immediately in the Schmidt canonical form
analogous to (7). The numerical
coefficients are the same. The state
transformers of ideal measurement are the
characteristic projectors $Q_2^k$ of the
measured observable $B_2$ \cite{Lahti95}.
Hence, the counterpart of $\ket{\psi
}_1^{fk}$ is evaluated from the analogues
of (4) and (7) utilizing (10):
$$ \ket{\Psi }_{12}^f\quad \rightarrow
\quad p_k^{(-1/2)}Q_2^k\ket{\Psi
}_{12}^f=\ket{\psi }_1^{fk}\otimes
\ket{\phi }_2^k.$$ Thus, the claimed
Schmidt canonical form (15) is obtained
with state vectors $\ket{\phi }_3^k$
satisfying the analogue of (10).\hfill
$\Box$

Expansion (15) implies that, as far as
subsystem $(1+2)$ is concerned, the two
measurements cause the following chain of
changes of state:
$$\ket{\psi }_1\otimes \ket{\phi }_2^0
\quad \rightarrow \quad \ket{\Psi
}_{12}^f\quad \rightarrow$$ $$
\rho_{12}^{ff}\equiv
\sum_k\Big(p_k\ket{\psi }_1^{fk}\bra{\psi
}_1^{fk}\otimes \ket{\phi }_2^k\bra{\phi
}_2^k\Big).$$ Neither $A_1$ nor $B_2$ are
incompatible with $\rho_{12}^{ff}$, i.
e., the incompatibility has disappeared
in the subsystem $(1+2)$. But it
reappears as the (same) amount of
incompatibility of $A_1$ with $\ket{\Phi
}_{123}^{ff}$ or rather with the
correlations between subsystem $1$ and
subsystem $(2+3)$.

Since $\ket{\Phi }_{123}^{ff}$ is
symmetric under any permutation of the
indices $1,2,3$, one can also say that in
this state $B_2$ is incompatible with the
correlations of subsystem $2$ with
subsystem $(1+3)$.

{\it In conclusion}, one can sum up the
results of this note as follows: In
quantum measurement of a discrete
observable $A$ in a quantum state
$\ket{\psi}$ two quantum-information
theoretic entities are of interest: One
of them is the amount of information that
is contained in $\ket{\psi}$ on $A$, and
the second is the {\it amount of
incompatibility} between $A$ and
$\ket{\psi}\bra{\psi}$. Both are
essentially preserved in the process of
predictive measurement of $A$ in
$\ket{\psi}$. The first entity reappears
as the quasi-classical part of the von
Neumann mutual information $I_{12}$ in
$\ket{\Psi}_{12}^f$, which is the final
state of the interaction of object and
measuring instrument. The second entity
reappears as the second part of $I_{12}$,
i. e., as the amount of entanglement in
$\ket{\Psi}_{12}^f$. Thus, {\it the
meaning of entanglement in repeatable
measurement is this (preserved) amount of
incompatibility}. One should keep in mind
that both entanglement and
incompatibility of observable and state
have no classical analogues, i. e., they
are purely quantum-mechanical concepts.

The result of this note is obtained under
the restriction of a pure initial state
of the object. It seems reasonable to
{\it conjecture} that it can be obtained
also for mixed initial states. An attempt
to confirm this conjecture will be
presented in a follow-up.

\end{document}